\documentclass[aps,pra,superscriptaddress,10pt]{revtex4-2}
\usepackage{amssymb,amsmath,amsthm,mathrsfs,amsfonts,dsfont}
\usepackage{bm}
\usepackage{physics}
\usepackage{braket}

\usepackage[dvipdfmx]{graphicx}
\usepackage{subfig}  

\usepackage[T1]{fontenc}  
\usepackage{newtxtext,newtxmath}  

\usepackage{color}
\usepackage{ulem}

\usepackage{enumerate}
\usepackage{here}
\usepackage{comment}
\usepackage{hyperref}

\hypersetup{colorlinks=true,citecolor=magenta,linkcolor=magenta,urlcolor=magenta}

\newcommand{\YM}[1]{\textcolor[rgb]{0.1, 0.5, 0.1}{#1}}

\begin{document}


\title{Linear Regression Using Quantum Annealing with Continuous Variables}


\author{Asuka Koura}
\email[These authors equally contributed to this paper.]{}
\affiliation{Department of Electrical, Electronic, and Communication Engineering, Chuo University, 1-13-27 Kasuga, Bunkyo-ku, Tokyo, 112-8551 JAPAN}
\affiliation{Global Research and Development Center for Business by Quantum-AI Technology (G-QuAT), AIST, Central2, 1-1-1Umezono, Tsukuba, Ibaraki 305-8568, JAPAN}

\author{Takashi Imoto}
\email[These authors equally contributed to this paper.]{}
\affiliation{Global Research and Development Center for Business by Quantum-AI Technology (G-QuAT), AIST, Central2, 1-1-1Umezono, Tsukuba, Ibaraki 305-8568, JAPAN}

\author{Katsuki Ura}
\affiliation{Global Research and Development Center for Business by Quantum-AI Technology (G-QuAT), AIST, Central2, 1-1-1Umezono, Tsukuba, Ibaraki 305-8568, JAPAN}

\author{Yuichiro Matsuzaki}
\email[]{ymatsuzaki872@g.chuo-u.ac.jp}
\affiliation{Department of Electrical, Electronic, and Communication Engineering, Chuo University, 1-13-27 Kasuga, Bunkyo-ku, Tokyo, 112-8551 JAPAN}


\begin{abstract}
Linear regression is a data analysis technique, which is
categorized as supervised learning. By utilizing known
data, we can predict unknown data. Recently, researchers
have explored the use of quantum annealing (QA) to perform linear regression where parameters are approximated to discrete values using binary numbers. However,
this approach has a limitation: we need to increase the
number of qubits to improve the accuracy. Here, we pro-
pose a novel linear regression method using QA that leverages continuous variables. In particular, the boson system facilitates the optimization of linear regression without resorting to discrete approximations, as it directly manages continuous variables while engaging in QA. 
The major benefit of our new approach is that it can ensure accuracy without increasing the number of qubits as long as the adiabatic condition is satisfied.
\end{abstract}
 
\maketitle


\date{\today}



%
\section{Introduction}
\label{sec:intro}

Quantum annealing (QA)\cite{kadowaki1998quantum,apolloni1989quantum,finnila1994quantum,farhi2000quantum,farhi2001quantum} is a technique for solving combinatorial optimization problems.
The combinatorial optimization problem can be transformed into finding the ground state of the Ising model. 
QA starts by preparing the system in the ground state of a Hamiltonian known as the Driver Hamiltonian, which has a simple ground state. The system is then gradually transformed from the Driver Hamiltonian to the Ising Hamiltonian (the Problem Hamiltonian), whose ground state represents the solution that we are exploring. 
This transformation is controlled by a time-dependent Hamiltonian.

In general, it is known that we can obtain the ground state
with high probability as long as the dynamics is adiabatic. This is guaranteed
by the adiabatic theorem \cite{kato1950adiabatic,jansen2007bounds,ambainis2004elementary}.
However, if the first-order phase transfition occurs during the time evolution, the energy gap becomes exponentially small, and it takes an exponentially long time to satisfy the adiabatic condition. To avoid such difficulties, the addition of a non-stoquastic term has been proposed. 

For some models, it has been reported that this addition not only avoids the first-order phase transition but also leads to a second-order phase transition 
\cite{seki2012quantum,seki2015quantum,susa2023nonstoquastic, takada2020mean}.
Additionally, it has been shown that by using a non-stoquastic term and auxiliary qubits, QA can perform computations equivalent to gate-type quantum computation\cite{biamonte2008realizable,mizel2007simple,aharonov2008adiabatic}.

There are also various applications of QA, such as quantum chemistry \cite{babbush2014adiabatic}, cryptanalysis \cite{jiang2018quantum,joseph2020not,joseph2021two,ura2023analysis}, and machine learning \cite{benedetti2017quantum, kurihara2014quantum,kumar2018quantum}. Particularly in the efficiency of the training part of machine learning, linear regression, SVM, \ and neural networks have been studied for various problems in recent years \cite{date2021adiabatic,chen2023quantum,pudenz2013quantum,dema2020support,bosch2023neural}.
To use QA for machine learning, we need to replace the cost functions with the Ising type Hamiltonian. In this case, the continuous variable should be represented with the discrete variables due to the discrete nature of the qubits.
There are various methods, including binary encoding and one-hot encoding \cite{chancellor2019domain,karimi2019practical,tamura2021performance}. However, many qubits are needed to obtain a highly accurate solution because conventional methods approximate continuous-valued parameters with discrete values.

We propose a method to encode continuous variables more efficiently than conventional methods for linear regression training in QA by using bosonic systems.
In the case of bosonic systems, each mode can provide an infinitely large Hilbert space. Therefore, our proposed QA can directly handle continuous values by constructing a problem Hamiltonian in the modes of the boson system. Unlike conventional linear regression using QA with qubits, our method allows us to obtain a highly accurate solution by setting a long annealing time. Numerical calculations have shown that adding a non-linear term 
during the QA process can lead to a more efficient solution for some cases.
It is worth mentioning that, in our previous work, we analyzed the case for a smaller number of bosonic systems\cite{kouraSSDM}.
On the other hand, in this paper, we investigate the case when we increase the number of cavities up to three.

This paper is organized as follows. Sections 2 and 3 review QA and linear regression.
Section 4 reviews linear regression with conventional QA.
Section 5 describes the method for solving linear regression by encoding the cost function with continuous variables. Section 6 presents the numerical results for evaluating the performance of our proposed method. Finally, section 7 presents the conclusions.

\section{Quantum annealing (QA)}
In this section, we review the quantum annealing (QA)
\cite{kadowaki1998quantum,apolloni1989quantum,finnila1994quantum,farhi2000quantum,farhi2001quantum}.
QA is one of the methods to solve the combinational optimization problem.
First, we introduce the total Hamiltonian as follows

\begin{align}
     \hat{H}(t)=\frac{t}{T}\hat{H}_p+\left(1-\frac{t}{T}\right)\hat{H}_d
    \label{qatotal}
\end{align}
where $t$ is the time, $T$ is the annealing time, $H_d$ is the driver Hamiltonian whose ground state is trivial, and $\hat{H}_p$ is the
problem Hamiltonian whose ground state corresponds to the solution.
In many cases, the transverse field
\begin{align}
    \hat{H}_d = -\sum_i \hat{\sigma}_i^x.
\end{align}
is chosen as a drive Hamiltonian.
QA consists of the following three steps.
First, prepare the ground state of $\hat{H}_d$ as the initial state at $t=0$. Second, let the system slowly evolve by $\hat{H}(t)$. Finally, at $t=T$, we obtain the ground state of $\hat{H}_p$ if the dynamics are adiabatic.

For the adiabatic dynamics, the following adiabatic conditions should be satisfied
\begin{align}
    \frac{|\braket{E_{1}(t)|\partial_{t}{\hat H(t)}|E_{0}(t)}|}{(E_1(t)-E_0(t))^2}\ll 1
\end{align}
where $|E_0(t)\rangle$ ($|E_1(t)\rangle$) represents the ground (first excited) state at time $t$ and $E_0(t)$ ($E_1(t)$) represents the ground (first excited) energy at time $t$.

If the first-order phase transition occurs during the time evolution, the energy gap $(E_1(t)-E_0(t))$ becomes exponentially small at this point, making it difficult to track the ground state \cite{sachdev1999quantum}. 
It is therefore important to consider ways of avoiding first-order phase transitions.
If we adopt the ferromagnetic p-spin model as the problem Hamiltonian, the first-order phase transitions occur during QA
\cite{jorg2010energy}.
In this case by adding a non-stoquastic term described as follows

\begin{align}
     \hat{V}_{AFF}= +N\left(\frac{1}{N}\sum_i^{N} \hat{\sigma_i^x}\right)^2,
     \label{Nonstoquastic}    
\end{align}
to the Hamiltonian, the first-order phase transition is avoided 
, and
the second-order phase transition appears \cite{seki2012quantum}. 
The same phenomenon is observed in the Hopfield model \cite{seki2015quantum}. It is also known that universal quantum computation can be 
performed by adding a certain type of
non-stoquastic term to QA \cite{biamonte2008realizable,mizel2007simple,aharonov2008adiabatic}.

\section{Linear regression}
Linear regression is classified as supervised learning and is a widely used statistical machine learning technique.
There are two typical approaches in supervised learning.
One of them is regression, which predicts real values from given input variables. The other is classification, where we assign a label to an input after we learn labeled training data.

Simple regression predicts a single output variable from a single input variable, while multiple regression predicts a single output variable from multiple input variables. Since both methods are supervised learning, preparing input and output variables as training data is necessary.

The multiple regression analysis is described here.
Let us define the linear model for the parameters as follows

\begin{align}
        f_{\boldsymbol{\theta}} (\mathbf{x})&= 
        \theta_1\phi_1(x_1) + \theta_2\phi_2(x_2) + \ldots + \theta_M\phi_M(x_M)\nonumber \\
          &=\sum_{m=1}^{M} \theta_m \phi_m(x_m)\nonumber\\
          &=\boldsymbol{\theta}^T\boldsymbol{\phi}(\mathbf x)\label{eq:lieanr_regression_model}
\end{align}
where $\{\theta_m
\}_{m=1}^M$ are the parameters, 
$\{\phi_m(x_m)
\}_{m=1}^M$ are the basis functions,
and $\{x_m
\}_{m=1}^M$ are the input variables.

In this study, the training data is represented by $\{\mathbf{x_i}, y_i\}_{i=1}^{N}$ where $N$ is the number of data. 
Here,
$\mathbf{x_i}$ is the input variable represented as $\mathbf{x_i}=(x_{i1},x_{i2},\ldots,x_{iM})$ where $M$ is the number of parameters.
On the other hand,
$y_{i}$ is the objective variable.

In multiple regression analysis, it is necessary to find optimal values for all elements of $\boldsymbol{\theta}$. The sum of squares error between the predicted value from this model and the target value is used as the cost function.
The cost function $L(\boldsymbol{\theta})$ is expressed as
\begin{align}
        L(\boldsymbol{\theta})=\frac{1}{2}\sum_{i=1}^{N}\left(f_{\boldsymbol{\theta}} (\mathbf{x}_i)-y_i\right)^2.
    \label{error}
\end{align}
We have

\begin{align}
    L(\boldsymbol{\theta})&=\frac{1}{2}\sum_{i=1}^{N}
        (-y_i+\sum_{m=1}^{M} \theta_m \phi_m(x_{im}))^2 \nonumber\\
        &=\frac{1}{2}(\sum_{i=1}^{N}
        \sum_{m,m'=1}^{M} \theta_m \theta_{m'} \phi_m(x_{im})\phi_{m'}(x_{im'}))
        -(\sum _{i=1}^N\sum_{m=1}^{M} \theta_m \phi_m(x_{im})y_i)+\frac{1}{2}(\sum _{i=1}^N y_i^2)
\end{align}


Let us define the following matrix.
\begin{align}
        \mathbf{\Phi}&=
        \boldsymbol{\phi}(\mathbf x)^T\\
        &=
            \begin{bmatrix}
            \phi_1(x_{11}) & \phi_2(x_{12}) & \ldots & \phi_m(x_{1M}) \\
            \phi_1(x_{21}) & \phi_2(x_{22}) & \ldots & \phi_m(x_{2M}) \\
            \vdots & \vdots & \vdots & \vdots \\
            \phi_1(x_{N1}) & \phi_2(x_{N2}) & \ldots & \phi_m(x_{NM}) \\
            \end{bmatrix}
            \label{cost}
\end{align}
We also define
\begin{align}
    \mathbf{y}=
    \begin{pmatrix}
 y_1 \\ y_2 \\ \cdot \\ \cdot \\ \cdot
 \\ y_N 
\end{pmatrix}
\end{align}
and
\begin{align}
        \mathbf{\theta}=
    \begin{pmatrix}
 \theta_1 \\ \theta_2 \\ \cdot \\ \cdot \\ \cdot
 \\ \theta_M 
\end{pmatrix}.
\end{align}
We can rewrite the cost function as follows.

\begin{align}
        L(\boldsymbol{\theta})=\frac{1}{2}(\theta ^{\rm{T}} \mathbf{\Phi}^{\rm{T}}\mathbf{\Phi}\theta 
        )-(
        \mathbf{\theta}^{\rm{T}}\mathbf{\Phi}^{\rm{T}}\mathbf{y})
        +\frac{1}{2} (\mathbf{y}^{\rm{T}}\mathbf{y}
        ).\label{eq:discrete_cost_function}
\end{align}

\section{Linear regression using conventional quantum annealing}

Here, we review the conventional training method for linear regression with QA using qubits.
Quadratic Unconstrained Binary Optimization (QUBO) formulation is commonly used for QA\cite{date2021adiabatic}. Here, the purpose is to minimize a quadratic function of binary variables.
To use QUBO formulation for machine learning, we need to represent the machine learning cost functions with the binary variables that take either 0 or 1.
The parameter $\theta_{m}$, which is a continuous value, is approximately represented by binary expansion as follows:
\begin{align}
    \theta_{m}\YM{\simeq} \sum_{k=1}^{K}p_{k}\hat{\theta}_{mk}
\end{align}
where the accuracy vector is $\mathbf{P} = [p_{1}, p_{2}, \cdots, p_{K}]^T$. Each component of the accuracy vector is an integer power of 2. Also, $K$ is the number of precision levels, and $\hat{\theta}_{mk}$ is a binary decision variable indicating whether an entry in $\mathbf{P}$ is to be selected or ignored. By replacing $\hat{\theta}_{mk}$ with $(\hat{1}+\hat{\sigma}^{(m,k)}_z)/2$, the cost function can be converted to QUBO.
In this case, the cost function can be represented by the Ising type Hamiltonian, and we can find its ground state with QA.
However,
the value of $\theta_m$ obtained by this formulation is an approximation of a continuous value by a discrete value. This means that,
to improve its accuracy, it is necessary to increase the number of qubits $K$ assigned to each parameter. 
Since the number of parameters is $M$, $K \times M$ qubits are required.
If the number of qubits available for QA is limited, we cannot obtain an accurate solution due to the approximation even if the adiabatic conditions are satisfied.

\section{Linear regression with continuous variables} 
Here, we explain our method to perform the linear regression with QA by using a bosonic system.
Let us transform the linear regression cost function
\ref{eq:discrete_cost_function} as follows
\begin{align}
        L(\boldsymbol{\theta}) &=\bf{\boldsymbol{\theta}^T\Phi^T\Phi\boldsymbol{\theta}}-2\mathbf{\boldsymbol{\theta}^T\Phi^Ty}\nonumber\\
        &=\sum_{m \neq l}\theta_m A_{ml} \theta_l + \sum_{m}A_{mm} \theta_m^2 -\sum_{m}B_m \theta_m
\end{align}
where $A=\mathbf{\Phi^T\Phi}$ and $B=2\mathbf{\Phi^Ty}$.
In this form, the equation includes both quadratic and linear terms in the parameters, while constant terms, which do not affect the result, are ignored.

The optimal parameters can be found by searching for a combination of variables $\theta_m$ to minimize the cost function. 
Let us perform the following replacement
\begin{align}
        \theta_m^2 &\rightarrow \hat{a}_m^\dagger \hat{a}_m,\\ \quad \theta_m &\rightarrow \frac{\hat{a}_m + \hat{a}_m^\dagger}{2}, \quad \\
        \theta_m\theta_l &\rightarrow \frac{\hat{a}_m \hat{a}_l^\dagger}{2} + \frac{\hat{a}_m^\dagger \hat{a}_l}{2}.
\end{align}
where $\hat{a}_m$ ($\hat{a}^\dagger _m$) denotes an annihilation (creation) operator of the $m$-th bosonic mode.

We define the Hamiltonian after this replacement as the problem Hamiltonian. Since this Hamiltonian is quadratic with respect to the operators, the ground state is the coherent state, which is the eigenstate of the annihilation operator. So, the parameter $\theta_m$ corresponds to the amplitude of the coherent state.

The driver Hamiltonian is also chosen as follows. 
\begin{align}
        \hat{H}_d =\sum _{m=1}^M \omega _D\hat{a}_m^\dagger \hat{a}_m
\end{align}
Throughout this paper, we fix $\omega _D=1$.
Since the driver Hamiltonian and the problem Hamiltonian are defined, we can proceed with QA. 
After performing QA, we can prepare the ground state of the problem Hamiltonian as long as the adiabatic condition is satisfied. 
Then, we can measure the expectation value of $ \frac{\hat{a}_m + \hat{a}_m^\dagger}{2}$, which corresponds to $\theta_m$. 
Since the amplitude of the coherent state is a continuous value, we can directly treat the continuous variable.
This differs from conventional methods, where discrete variables approximate continuous values. 

We also consider adding the following non-linear terms to the annealing Hamiltonian as non-stoquastic catalysts.
\begin{align}
    \hat{H}_{ns}= \sum _{i=1}^N\chi  \left(\hat{a}_i^\dagger \hat{a}_i\right)^2 \label{nonlin} 
\end{align}
where
$\chi$ is called the Kerr nonlinearity.
The Holstein-Primakov transformation is a way to convert an operator of a spin ensemble into that of the bosonic system \cite{primakoff1939many}.

By using this transformation on the Eq.
\eqref{Nonstoquastic}, we obtain the Eq. \eqref{nonlin}. 

The total Hamiltonian for QA with the non-linear terms is described as follows.\cite{miyazaki2022effective}
\begin{align}
    \hat{H}(t) = \left(1 - \frac{t}{T}\right)\hat{H}_d + \frac{t}{T}\hat{H}_p + \left(1 - \frac{t}{T}\right)\frac{t}{T}\hat{H}_{ns}.
\end{align}

In this process Here, we start from the driver Hamiltonian and finish with the problem Hamiltonian, similar to the conventional QA. 
Additionally, we increase the strength of the non-linear terms during QA. This may improve the performance of our proposed method, which we will describe later.

\section{Numerical results}
\begin{figure}[h!t]
  \includegraphics[width=20cm]{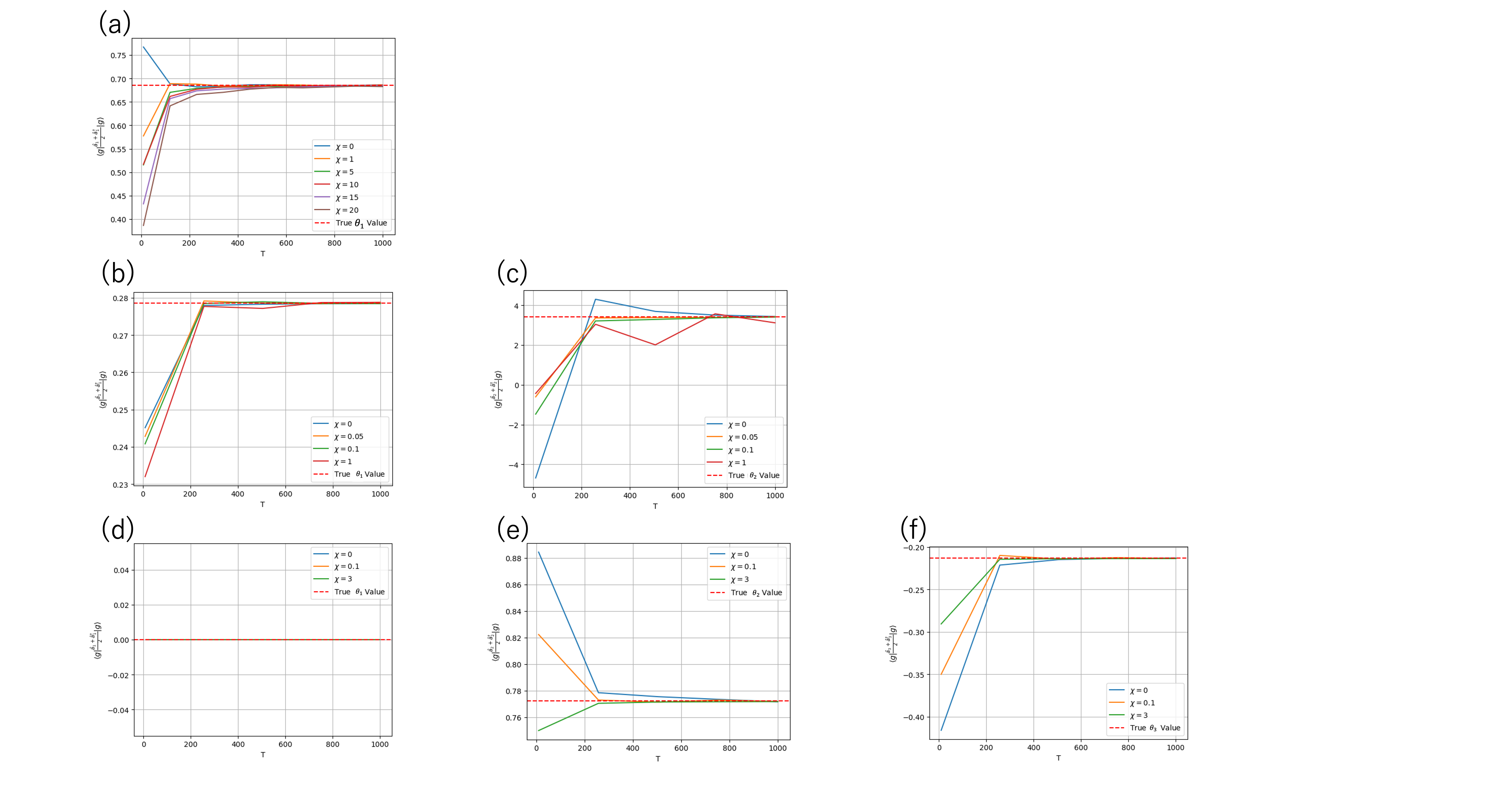}
  \centering
    \caption{Numerical results for linear regression with QA with bosonic systems are presented, where the expectation values of $\frac{\hat{a}_m + \hat{a}_m^{\dagger}}{2}$ in the state after QA 
    are plotted against $T$.
    (a) For $M=1$. The model is set as $f_\theta(x) = \theta_1 x$ where $\theta_1$ is the parameter.
    (b),(c) For $M=2$, the model is set as $f_{\boldsymbol{\theta}}(x) = \theta_1 x + \theta_2$ where $\theta_1$ and $\theta_2$ are the parameters.
    (d),(e),(f) For $M=3$, the model is set as $f_{\boldsymbol{\theta}}(x) = \theta_1 x_1 + \theta_2 x_2 + \theta_3$ where $\theta_1$, $\theta_2$, and $\theta_3$ are the parameters. As we increase $T$, the expectation values approach the true value.
    }
  \label{para_ex}
\end{figure}
In this section, we show our numerical results to evaluate the performance of our proposed method.
We choose the number of parameters as $M=1,2,3$.
For $M$ = 1, 2, we use
the sepal length and sepal width of setosa for the training where we obtain the
data from the Iris dataset \cite{fisher1936use}. 
The scikit-learn library is used for training on a classical computer \cite{scikit-learn}.
For $M$ = 3, \YM{we use}
the palmer penguin dataset \cite{horst2020palmerpenguins} 
for training.
Here, we define the body mass (g) as $x_1$, bill depth (mm) as $x_2$, and flipper length (mm) as $y$.
The same scikit-learn library is used to obtain the parameters with a classical computer \cite{scikit-learn}. 
In Figure \ref{para_ex}, we plot the expectation values of $\frac{\hat{a}_m+\hat{a}_m^{\dagger}}{2}$ in the states after QA against the annealing time $T$ for total parameters of $M$ = 1,2,3.
The dotted lines are the values of the parameters optimized by the classical computer.
By increasing $T$ for any number of parameters, the expectation value s of $\frac{\hat{a}+\hat{a}^{\dagger}}{2}$ approaches the values calculated by the classical computer.
Therefore, regardless of the number of parameters,
the correct solution can be obtained if the annealing time is sufficiently long.
This outcome is in agreement with the predictions of the adiabatic condition.

\begin{figure}[h!t]
  \includegraphics[width=20cm]{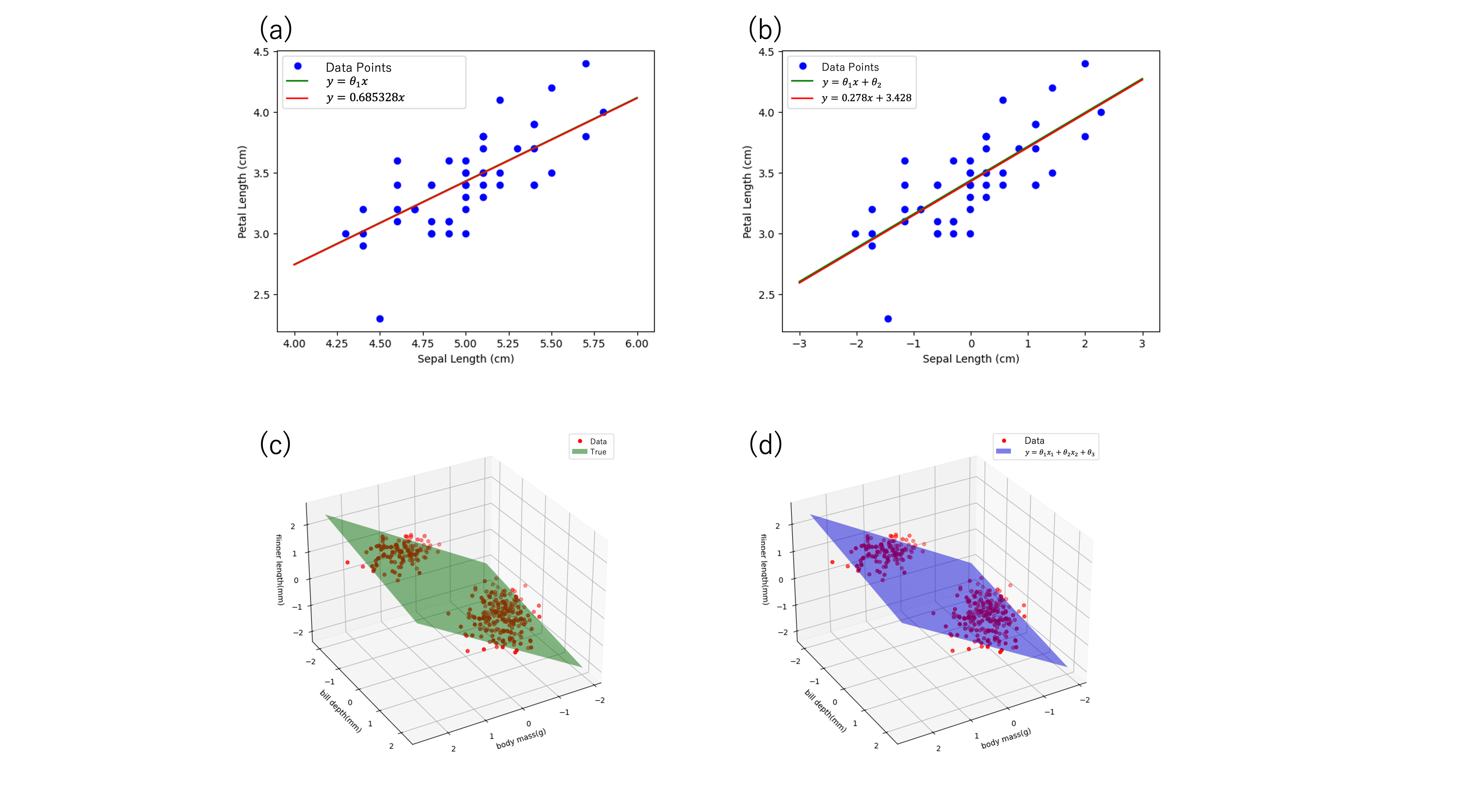}
  \centering
  \caption{
      Plot of the obtained model $f_\theta(x)$ with data. In our method, the expectation value of $\frac{\hat{a}_m+\hat{a}_m^{\dagger}}{2}$ obtained at $T=1000,\chi=0$ after QA is used as a model parameter. 
      (a) Plot of $f_\theta(x)=\theta_1x$ for $M=1$. Using the Iris dataset, scikit-learn library is used for the parameters with a classical computer \cite{scikit-learn}. For our method, we use the expectation value for the parameter. The vertical axis represents the sepal length and the horizontal axis represents the sepal width. (b) Plot of $f_\theta(x)=\theta_1x+\theta_2$ for $M=2$. Using the Iris dataset, scikit-learn library is used for the parameters with a classical computer, while we use our method with QA for the parameters. The vertical axis represents the sepal length, while the horizontal axis represents the sepal width. (c),(d) Plots of $f_\theta(x)=\theta_1x_1+\theta_2x_2+\theta_3$ for $M=3$. We adopt palmer penguin dataset \cite{horst2020palmerpenguins} using the parameters obtained by the classical computer method (c) and that obtained by our method (d), where the x-axis is body mass(g), the y-axis is bill depth(mm) and the z-axis is flipper length(mm).}
  \label{plot}
\end{figure}

\begin{figure}[h!t]
  \centering
  \includegraphics[width=18cm]{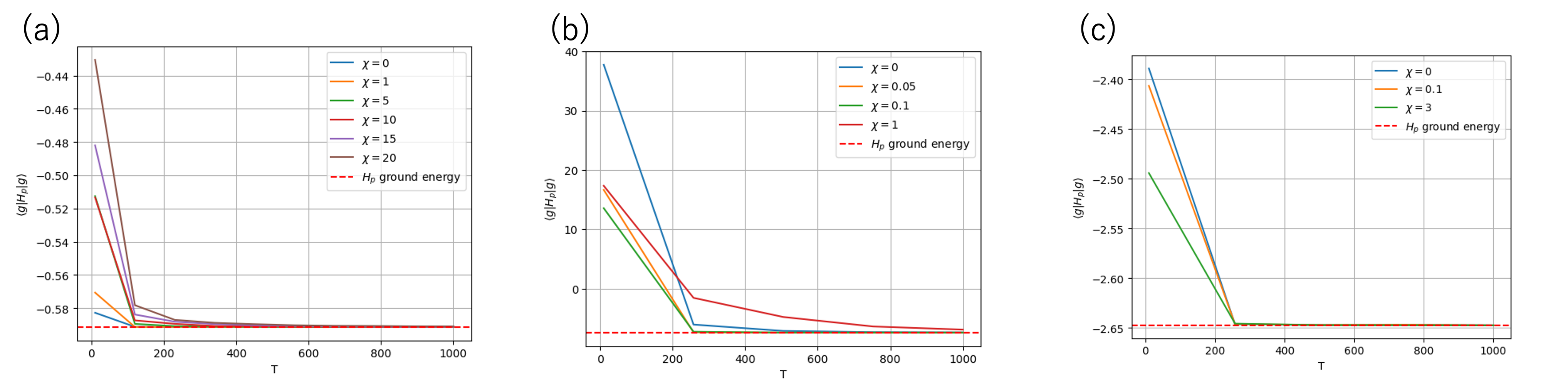}
  \caption{
      Plot of
      the expectation value of $\hat{H}_p$ of the state after QA
      against $T$.
    We plot the results of (a) $M=1$, (b) $M=2$, and (c) $M=3$. As we increase an annealing time $T$, the expectation value approaches the ground-state energy, which is consistent with the adiabatic theorem. Moreover,as we increase the non-linearity $\chi$, the expectation value becomes closer to the ground-state energy for a small annealing time $T$.
         }
  \label{hp_ex}
\end{figure}

In Figure \ref{plot}, we also plots the data points and the predicted value using $f_\theta(x)$ Eq.\ref{eq:lieanr_regression_model} with the parameter values obtained from numerical results. The plot obtained from our method is similar to what obtained from the classical computation for $M=1,2,3$.
In Figure \ref{hp_ex}, we plot the expectation value of $\hat{H}_p$ after QA with $M=1,2,3$. 
The dotted line is the ground-state energy calculated by the exact diagonalization.
As we increase the annealing time $T$, the expectation value of $\hat{H}_p$ after QA becomes closer to the ground-state energy, which is consistent with the adiabatic condition.
Moreover, for a small $T$, the expectation value of $\hat{H}_p$ after QA approaches the ground-state energy as we increase $\chi$. This shows a potential that the non-linear terms may improve the performance of QA, although further research is needed
to assess the general utility for future work.

\section{Conclusion}
In conclusion, we propose linear regression with quantum annealing by using a bosonic system such as optical cavities.

In linear regression, it is necessary to optimize the continuous parameters that minimize the cost function.
We construct the Hamiltonian of the bosonic systems where the ground state is the coherent state, and the amplitude of the coherent state corresponds to the parameters of the target cost function. After we perform QA, we measure the amplitude of the coherent state, giving us the optimized parameter so that the cost function should be minimized. The adiabatic theorem guarantees that we can obtain the  solution with high accuracy as long as the dynamics is aidabatic.

The conventional QA for linear regression where discrete variables approximate the continuous variables, which requires many qubits. On the other hand, in our proposed method, the number of cavities only depends on the number of parameters in the cost function.
Therefore, we can expect a reduction in the number of quantum devices required for the experiment compared to the conventional method.
We also confirm that by adding a non-linear term to the Hamiltonian during QA, the expectation value of $\hat{H}_p$ after QA can approach the ground energy with a relatively short annealing time. This shows the potential of the non-linear terms to improve our method's performance. 
Our results enable bosonic quantum annealing (QA) for several types of machine learning with continuous variables.
We use the Qutip for numerical results\cite{johansson2012qutip}.

\section{acknowledgements}
This paper is partly
based on results obtained from a project, JPNP16007,
commissioned by the New Energy and Industrial Technology Development Organization (NEDO), Japan.
This work was supported by
JST Moonshot (Grant Number JPMJMS226C). Y. Matsuzaki is supported by JSPS KAKENHI (Grant Number
23H04390). This work was also supported by CREST
(JPMJCR23I5), JST.

\bibliography{ref}

\end{document}